\documentclass[twocolumn,aps,pra]{revtex4-1}

\usepackage{graphicx}
\usepackage{graphics}
\usepackage{dcolumn}
\usepackage{bm}
\usepackage{amsmath}
\usepackage{amssymb}
\usepackage{color}
\usepackage{float}
\usepackage{soul}
\begin{document}

\title{Coherence Protection of Electron Spin in Earth-field Range by All-optical Dynamic Decoupling}

\author{Peiyu Yang $^{1}$ }
\author{Guzhi Bao $^{2}$ }
\email{guzhi\_bao@126.com}
\author{L. Q. Chen $^{1,4}$ }
\email{lqchen@phy.ecnu.edu.cn}
\author{Weiping Zhang $^{2,3,4}$}
\email{wpz@sjtu.edu.cn}
\date{\today}
\affiliation{$^{1}$ State Key Laboratory of Precision Spectroscopy, Quantum Insitute for Light and Atoms, School of Physics and Electronic Science,
	East China Normal University, Shanghai 200241, China. \\
$^{2}$ School of Physics and Astronomy, and Tsung-Dao Lee Institute, Shanghai Jiao Tong University, Shanghai 200240, China.\\
$^{3}$ Collaborative Innovation Center of Extreme Optics, Shanxi University, Taiyuan, Shanxi 030006, China.\\
$^{4}$ Shanghai Research Center for Quantum Sciences, Shanghai 201315, China.}

\begin{abstract}
In recent years, unshielded atomic systems have been attracting researchers’ attention, in which decoherence is one of the major
problems, especially for high precision measurements. The nonlinear Zeeman effect and magnetic field gradient are the main decoherence sources of atomic electron spin in Earth-field range. Here, we propose a method to cancel out the two dominant broadening effects simultaneously by an all-optical dynamic decoupling approach based on Raman scattering in the $^{87}{\rm Rb}$ Zeeman sublevels. By adjusting the parameters of the Raman lasers, we realize spin control along an arbitrary direction. We analyze the state evolution of atomic spin under the Raman light control sequence in detail. The results show that both the nonlinear Zeeman effect and magnetic field gradient can be significantly suppressed.
\end{abstract}

\maketitle
\section{Introduction}
Atomic electron spin is a basic information carrier in quantum science  \cite{kominis2003subfemtotesla,fixler2007atom,dunning2014composite,chen2015atom,bao2018suppression,bao2018all,fu2020sensitive,saywell2020optimal,jiang2020interference,dudin2013light,chen2010observation,rui2015operating,guo2019high,brennen2005criteria,wu2016realization}. Coherence protection is of major importance for precision measurements and for scalable quantum information networks.  Short coherence time limits the sensitivity of atomic sensors\cite{kominis2003subfemtotesla,fixler2007atom,dunning2014composite,chen2015atom,bao2018suppression,bao2018all,fu2020sensitive,saywell2020optimal} and quantum memory time\cite{dudin2013light,rui2015operating,guo2019high}. Such as, the sensitivity of atomic magnetometer is described as $ \delta B=1/(\gamma\sqrt{nT_2 Vt}) $\cite{kominis2003subfemtotesla}, in which $ T_2 $ represents the coherence time. Longer coherence time indicates better sensitivity. Usually, high-performance atomic systems are protected by magnetic shields \cite{kominis2003subfemtotesla,chen2010observation,chen2015atom,guo2019high,dudin2013light}. However, practical application often requires that atomic systems work in challenging conditions where background magnetic fields with large amplitudes or large gradients exist \cite{fu2020sensitive}. When an atomic sample operates in the Earth's magnetic field, the nonlinear Zeeman ($\textrm{NLZ}$) effect and magnetic field gradient are major decoherence sources of atomic electron spins \cite{acosta2006nonlinear,seltzer2007synchronous,jensen2009cancellation,fu2020sensitive,bao2018suppression,bao2018all}. These decoherence effects broaden the linewidth of magnetic resonance and decrease the signal intensity, which leads to the drops of the sensitivity quadratically to the coherence time. Furthermore, the asymmetries of magnetic-resonance signal due to the NLZ splitting is particularly troublesome in airborne and marine systems \cite{alexandrov2003recent}. Several ways have been reported to cancel the $\textrm{NLZ}$ effect, e.g., double-modulated synchronous optical pumping \cite{seltzer2007synchronous}, high-order polarization moments \cite{acosta2006nonlinear}, tensor light shift effects \cite{jensen2009cancellation} and spin locking \cite{bao2018suppression}. Among them, spin locking is robust against orientation changes \cite{bao2018suppression}. However, spin locking brings power broadening to the magnetic resonance, limiting the sensitivity \cite{bao2018all}. To achieve a narrow linewidth, the power of spin locking needs to be finely controlled to reach a balance between spin locking and power broadening. With the exisitence of a tilted angle between the sensor and the leading field, the microwave fields and the atomic spins are no longer rotating in the same plane which decreases the effect of spin locking.

Another choice is dynamic decoupling ($\textrm{DD}$) \cite{dudin2013light,rui2015operating,antonijevic2003refocussing,de2010universal,naydenov2011dynamical,shaniv2019quadrupole,pedernales2020motional,Miaoeabc5186,hahn1950spin,barthel2010interlaced,ahmed2013robustness,pokharel2018demonstration,upadhyay2020ultralong}, which is a quantum control technique used to extend the atomic or nuclear coherence time by periodic sequences of instantaneous modulation pulses and has been widely used to cancel quadrupolar energy shifts \cite{antonijevic2003refocussing,shaniv2019quadrupole} and inhomogeneous broadenings \cite{hahn1950spin,de2010universal,rui2015operating}. Conventional $\textrm{DD}$ is realized by pulsed microwave fields \cite{dudin2013light,antonijevic2003refocussing,de2010universal,naydenov2011dynamical,shaniv2019quadrupole,pedernales2020motional,Miaoeabc5186}. However, global applied microwave fields might lead to crosstalk between closely located samples \cite{bao2018all,bison2009room,lembke2014optical,borna201720}. Additionally, in an atom interferometer, the microwave fields cannot generate a recoil velocity to separate the atomic energy levels \cite{fixler2007atom,dunning2014composite,saywell2020optimal}. Moreover, the microwave fields cannot be applied to a faraway atomic sample directly, which limits the applicability of this technique to remote control \cite{bao2018all,bustos2018remote,patton2012remotely}. An alternative solution is needed to overcome these drawbacks above.

We construct an all-optical pulsed control field to achieve spin rotation by Raman scattering. Raman scattering is a simple way to couple atomic spins with only two optical fields \cite{fixler2007atom,chen2010observation,dunning2014composite,rui2015operating,chen2015atom,guo2019high,saywell2020optimal}. Here, we manage to rotate the spin state by an arbitrary angle along any direction by adjusting the Rabi frequencies, detunings, propagating directions and pulse lengths of Raman lasers. By inserting $\textrm{Hahn}$ echo $\tau-180^{\circ}y-\tau$ into quadrupolar echo $\tau-90^{\circ}y-\tau$, we design an all-optical sequence to cancel the $\textrm{NLZ}$ effect and magnetic field gradient simultaneously in the Earth-field range. Since all the control fields are along $\hat{y}$ axis, even with a large tilted angle, DD works successfully. To evaluate the performance of this all-optical scheme, we exhibit the state evolution process and analyze the fidelity of the atomic spin state with and without $\textrm{DD}$. The results show that the fidelity up to $99.99\%$ with DD is much better than the free evolution case, indicating that the coherence of the atomic spin state is well maintained by DD. 

In this work, we propose and design an all-optical DD sequence by Raman process to prolong the coherence time of atomic system in the Earth-field range. The article is organised as follows. In Sec. II we study the underlying dephasing mechanism limiting the coherence time. In Sec. III we give an effective rotation Hamiltonian by Raman process to manipulate the atomic spin in an all-optical way. In Sec. IV we design an all-optical DD sequence and analyze the DD performance. In the last section, we draw some conclusions and discussions.

\section{Dephasing mechanism of the electron spin}
The dephasing effects of the electron spin contain at least six components: spin-exchange relaxation, spin-destruction relaxation, wall-collision relaxation, optical power broadening, the $\textrm{NLZ}$ effect and magnetic field gradient broadening. Among them, the $\textrm{NLZ}$ effect and magnetic field gradient are dominant in most cases. 

The eigenvalues for the ground state $5 S_{1/2}$ of $^{87}\mathop{\rm Rb}$ in an external magnetic field B are given by the Breit-Rabi formula \cite{breit1931measurement,auzinsh2010optically} and illustrated in Fig.\,1. The energy splittings between adjacent Zeeman sublevels are
\begin{equation}
	E_{m_F+1}-E_{m_F}=\pm\frac{\mu_BB}{2}\mp\hbar \omega(2m_F+1),
\end{equation}
where $m_F$ is the magnetic quantum number, $\mu_B$ is the Bohr magneton, B is the magnetic field intensity, $\omega=(\mu_BB)^2/(4\hbar\delta)$ is the quantum-beat revival frequency caused by the $\textrm{NLZ}$ effect, and $\delta$ is the hyperfine-structure energy splitting; $\pm$ refer to the $F=I\pm1/2$ hyperfine components, and $I$ is the nuclear spin. In the $\textrm{Earth-field}$ range (50\,$\mu$T), the revival frequency for $^{87}\mathop{\rm Rb}$ is $\omega=2\pi\times17.9$\,Hz, which is comparable to the width of magnetic resonance. The right side of Fig.\,1 gives the split magnetic-resonance signal in the presence of the $\textrm{NLZ}$ effect.

\begin{figure}[htbp]
	\includegraphics[width=1\columnwidth]{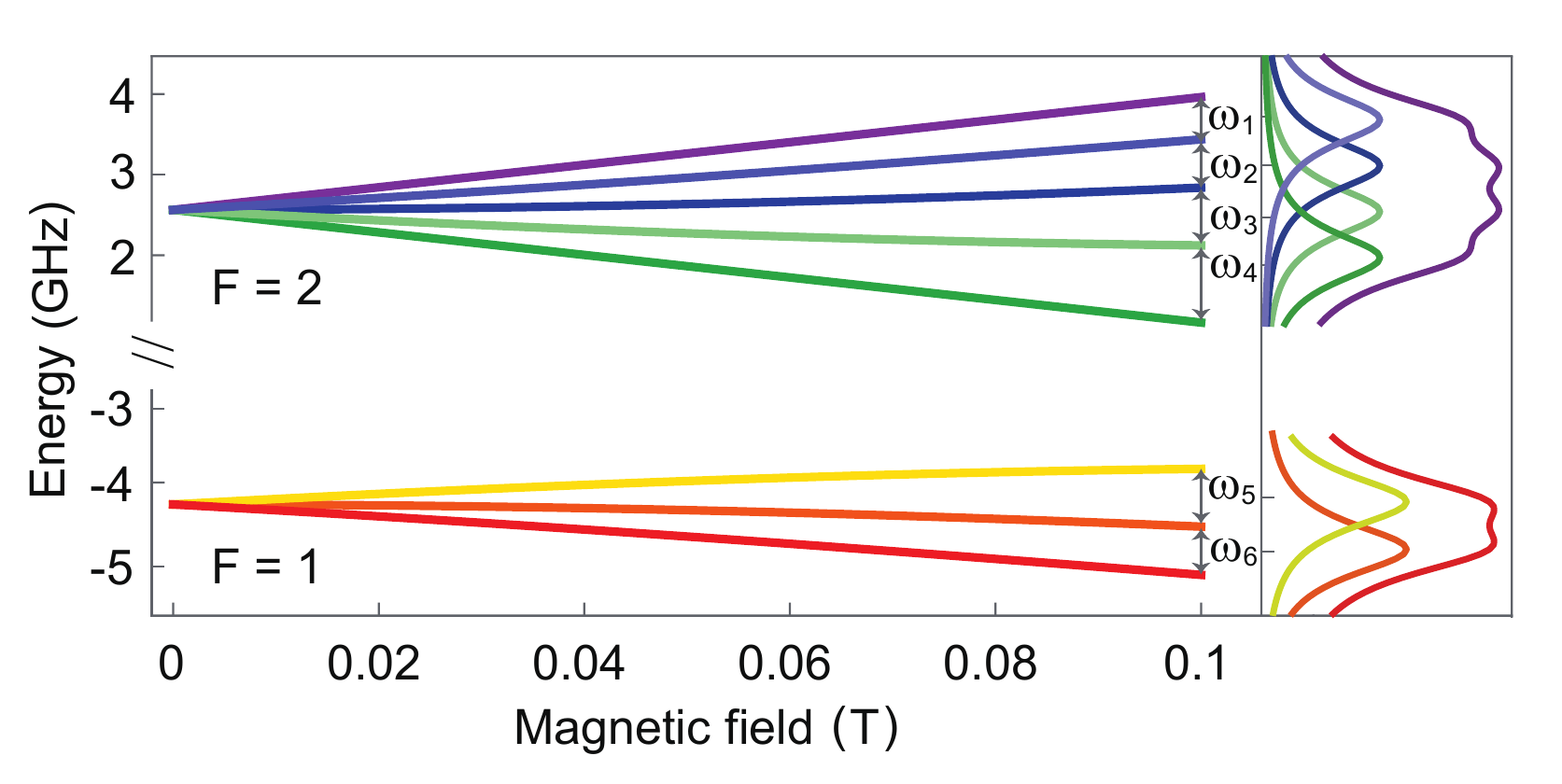} \caption{ Hyperfine structure of the $5S_{1/2}\,F=1$ and $F=2$ state manifolds of the $^{87}\mathop{\rm Rb}$ atom in an external magnetic field. The right side shows the split magnetic-resonance signal for the $F=2$ states (up) and $F=1$ states (down). The signals are fitted with Lorentzian peaks arising due to the $\textrm{NLZ}$ effect. $\omega_1= \Omega_L-2\omega$, $\omega_2=\Omega_L-\omega$, $\omega_3=\Omega_L+\omega$ and $\omega_4=\Omega_L+2\omega$ are the quantum beating frequencies of $F=2$.  $\omega_5=\Omega_L+\omega$ and $\omega_6=\Omega_L-\omega$ are the quantum beating frequencies of $F=1$.}  
\end{figure}

When atomic systems work in an open environment, without the protection of magnetic shields, background magnetic fields have large gradients. Due to the inhomogeneous distribution of the external magnetic field, atoms located in different positions have different Larmor precession frequencies, which leads to dephasing of the atomic electron spin.	

With consideration of the two main dephasing effects ($\textrm{NLZ}$ effect and magnetic field gradient), the total Hamiltonian of our system on the basis of the Zeeman sublevels with a magnetic field along the $\hat{z}$ axis is
\begin{equation}
	H=\hbar(\Omega_LJ_z+\Delta\Omega_LJ_z+\omega J_z^2),
\end{equation}
where $\hbar$ is Plank's constant, $J_z$ is the angular momentum along the $\hat{z}$ axis, $\Omega_L$ is the Larmor precession frequency, $\Delta\Omega_L$ is the inhomogeneous frequency broadening caused by the magnetic field gradient, and $\omega$ is the revival frequency due to the $\textrm{NLZ}$ effect. Our goal is to cancel out the $\textrm{NLZ}$ effect and magnetic field gradient simultaneously.

\section{Spin control by Raman processes}
A dynamic decoupling experiment often requires fast spin rotations by certain degrees in the $\hat{x}$ or $\hat{y}$ direction, denoted as $\Phi_x$ or $\Phi_y$, where $\Phi$ is the angle of rotation. Raman scattering is a typical method for manipulating the atomic electron spin. For the all-optical scheme, we construct an equivalent $J_x$ or $J_y$ Hamiltonian by quantum interference of Raman channels.	
\begin{figure}[htbp]
	\includegraphics[width=1\columnwidth]{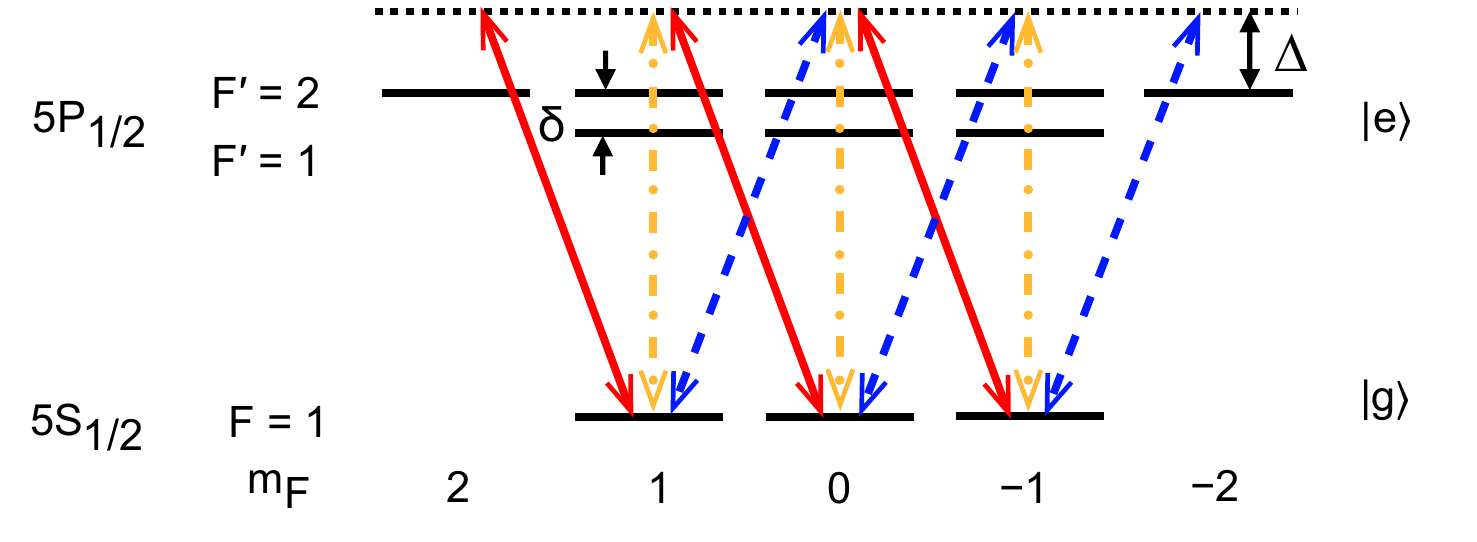} \caption{Transition channels of the $^{87}\mathop{\rm Rb}$ D1 line from the ground state $5S_{1/2}, F=1$ Zeeman sublevels to excited energy levels $5P_{1/2}, F'=1$ and $F'=2$. $\delta$ is the hyperfine energy level splitting. $\Delta$ denotes Raman detuning. $m_F$ is the magnetic quantum number. Solid lines represent $\sigma_+$ transitions. Dashed lines represent $\sigma_-$ transitions. Dot-dashed lines represent $\pi$ transitions.} 
\end{figure} 

Here, we consider the D1 transition of $^{87}\mathop{\rm Rb}$ from $5S_{1/2}, F=1$ to $5P_{1/2}, F'=1$ and $F'=2$. We apply laser lights with three polarizations, i.e., $\sigma_+$, $\pi$ and $\sigma_-$, the Rabi frequencies of which are $\Omega_1, \Omega_2$ and $\Omega_3$, respectively. The propagating directions of the Raman lasers are all along the $\hat{x}$ axis. Possible transition channels are depicted in Fig.\,2. The hyperfine energy level splitting of the excited state $\delta$ is 800\,MHz and the natural linewidth of the excited state is around 6\,MHz. When the Raman detuning $\Delta$ is far enough, the effective Hamiltonian $H_{\rm eff}$ on the basis of the Zeeman sublevels can be derived from the method in the Appendix A. To achieve a $J_x$-rotation Hamiltonian, we assume $\Omega_3=\Omega_1$ and that $\pi$ polarized light has a $\pi/2$ phase difference with circularly polarized lights $\sigma_+$, $\sigma_-$. We can finally write $H_{\rm eff}$ as		
\begin{equation}
	\begin{aligned}
		H_{\rm eff}&=
		\alpha_1\left(                 
		\begin{array}{ccc}   
			0 & 1 & 0\\  
			1 & 0 & 1\\ 
			0 & 1 & 0\\
		\end{array}
		\right)+\alpha_2\left(                 
		\begin{array}{ccc}   
			0 & 0 & 1\\  
			0 & 0 & 0\\ 
			1 & 0 & 0\\
		\end{array}
		\right)\\
		&+\beta_1\left(                 
		\begin{array}{ccc}   
			1 & 0 & 0\\  
			0 & 1 & 0\\ 
			0 & 0 & 1\\
		\end{array}
		\right)+\beta_2\left(                 
		\begin{array}{ccc}   
			1 & 0 & 0\\  
			0 & 0 & 0\\ 
			0 & 0 & 1\\
		\end{array}
		\right),
	\end{aligned}
\end{equation}
where $\alpha_1,\alpha_2$, $\beta_1$ and $\beta_2$ are the coupling coefficients, with
\begin{subequations} \label{e:1234}
	\begin{align}
		\label{12345}	
		\alpha_1&=\frac{(7\delta+4\Delta)\Omega_1\Omega_2}{6(3\delta-4\Delta)(5\delta+4\Delta)},\\    
		\label{12345}     
		\alpha_2&=-\frac{\delta\Omega_1^2}{3(3\delta-4\Delta)(5\delta+4\Delta)},\\	 
		\label{12345}
		\beta_1&=-\frac{2(3\delta+4\Delta)\Omega_1^2+(5\delta+4\Delta)\Omega_2^2}{6(3\delta-4\Delta)(5\delta+4\Delta)},\\	 
		\label{12345}
		\beta_2&=-\frac{\delta_1(\Omega_1-\Omega_2)(\Omega_1+\Omega_2)}{3(3\delta-4\Delta)(5\delta+4\Delta)},
	\end{align}
\end{subequations}
where $\delta$ is the hyperfine energy level splitting and $\Delta$ is the Raman detuning.

The first term in Eq.\,(3) containing $\alpha_1$ denotes the $\Delta m_F=1$ transitions, which is a typical $J_x$ form for the $F=1$ system. The second term containing $\alpha_2$ represents transitions between the two outermost Zeeman energy levels $m_F=-1$ and $m_F=1$. The third term containing $\beta_1$ gives a scalar shift independent of the Zeeman sublevels. The scalar shift results in a global phase factor relative to the wavefunction, which has no influence on the density matrix. The last term proportional to $\beta_2$ is a tensor shift similar to the $\textrm{NLZ}$ shift. To obtain a clean $J_x$ Hamiltonian, we need to adjust the detunings and Rabi frequencies of the Raman fields to eliminate the second term and the last term. 

\begin{figure}[htbp]
	\includegraphics[width=1\columnwidth]{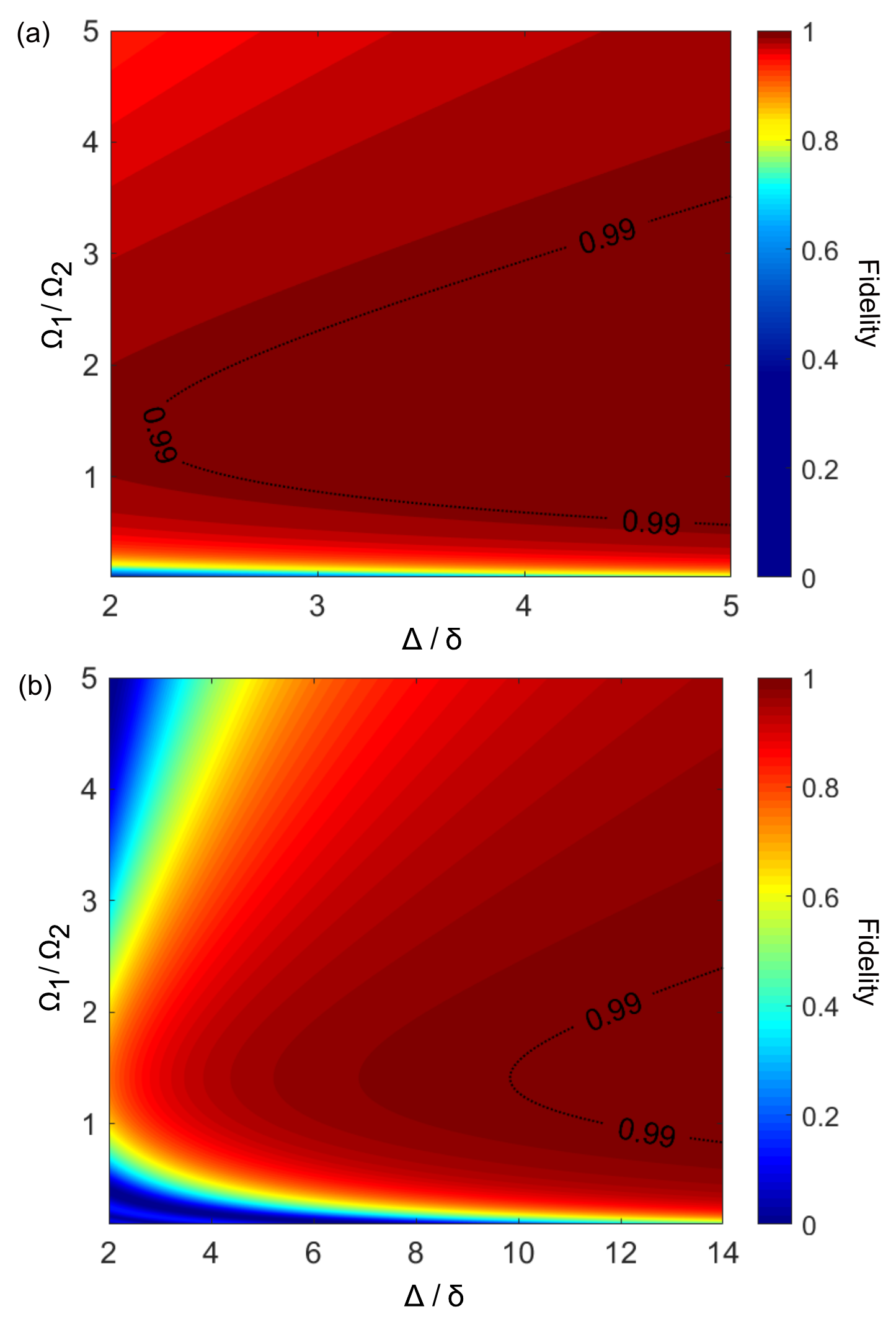} \caption{(a) The dependence of the fidelity on $\Omega_1/\Omega_2$ and $\Delta/\delta$ in the basis of $5S_{1/2}, F=1$. (b) The dependence of the fidelity on $\Omega_1/\Omega_2$ and $\Delta/\delta$ on the basis of $5S_{1/2}, F=2$. The red area where the fidelity is close to one indicates the optimized conditions to realize the $J_x$ Hamiltonian. The black dashed line is a contour line where the fidelity is 0.99. }
\end{figure} 

Since we want to obtain a pure $J_x$ Hamiltonian by Raman processes, we simulate how $J_z$ evolves with the application of $H_{\rm eff}$ in Eq.\,(3). $J_z$ is turned into $-J_y$ after a $90^{\circ}x$ pulse. To find the optimal detunings and Rabi frequencies of the Raman fields for a $J_x$ Hamiltonian, we calculate the fidelity of $-J_y$ and the state after a $90^{\circ}$ rotation operation by $H_{\rm eff}$. The fidelity formula is $\mathcal{F}= {\rm Tr}|\sqrt{\sqrt{\rho_{1}}\rho_{2}\sqrt{\rho_{1}}}|^2$ \cite{jeong2007quantum}. $ \rho_{1} $ is $ -J_y $ and $ \rho_{2} $ is the state after a $90^{\circ}$ rotation operation by $H_{\rm eff}$. If the fidelity is close to one, we can say that $H_{\rm eff}$ is a perfect $J_x$ Hamiltonian. 

Figure\,3a shows the dependence of the fidelity on the dimensionless parameters $\Omega_1/\Omega_2$ and $\Delta/\delta$ for $F=1$. When the Raman detuning is large enough ($\Delta \gg \delta$), the fidelity is close to one. Then, we can write the effective Hamiltonian as
\begin{equation}
	\begin{aligned}
		H_{\rm eff}=\frac{\Omega_1\Omega_2}{24\Delta}\left(    \begin{array}{ccc}   
			0 & 1 & 0\\  
			1 & 0 & 1\\ 
			0 & 1 & 0\\
		\end{array}
		\right).
	\end{aligned}
\end{equation}
The coefficient $\Omega_1\Omega_2/(24\Delta)$ is the effective Rabi frequency. 
\begin{figure*}[htpb]
	\includegraphics[width=2\columnwidth]{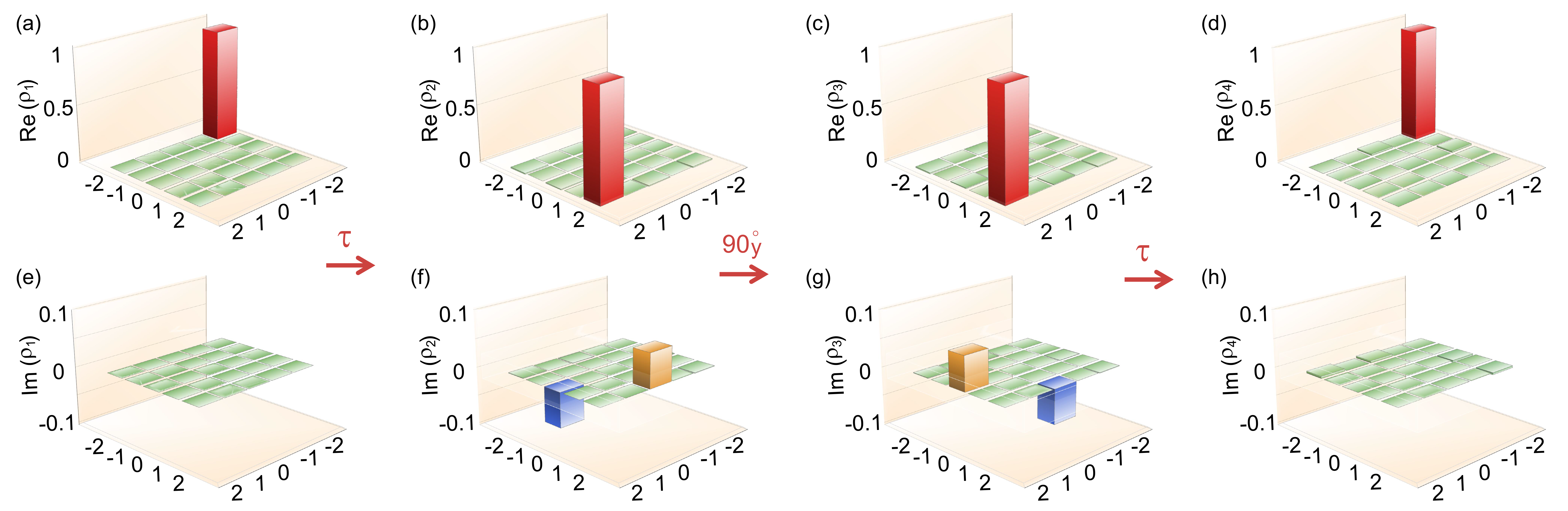} \caption{ The density matrix evolution under the quadrupolar echo $\tau-90^{\circ}y-\tau$ in the $J_y$ representation. (a-d) are real parts of the density matrix. (e-h) are imaginary parts of the density matrix. $\rho_1$ is the state $-J_y$. $\rho_2$ is the state after free evolution with the $\textrm{NLZ}$ dephasing effect. $\rho_3$ is the state after the $90^{\circ}y$ pulse. $\rho_4$ is the final state after $\textrm{DD}$ for a whole Larmor precession cycle.} 
\end{figure*}

Note that in Eq.\,(5), the coherence between $m_F=1$ and $m_F=-1$ disappears. This is caused by quantum interference between the two Raman transition channels from $5S_{1/2}, F=1$ to $5P_{1/2}, F'=1$ and $F'=2$. Under appropriate detunings and Rabi frequencies of the Raman lights, the two transition amplitudes form destructive interference.

Following the same steps, we can construct an efficient $J_x$ Hamiltonian for the energy level $5S_{1/2}, F=2$. Similarly, we can plot the fidelity of $-J_y$ and the state after a $90^{\circ}$ rotation operation by the $H_{\rm eff}$ on the basis of $5S_{1/2}, F=2$, as shown in Fig.\,3b. Compared with the $F=1$ case, the optimal condition for $F=2$ is more rigorous. We need a larger Raman detuning to achieve the same fidelity (dashed line in Fig.\,3). 

In total, there are three necessary conditions for Raman lights to construct a perfect $J_x$ rotation Hamiltonian: (i)\,The Rabi frequencies and phases of two circularly polarized lights are equal. (ii)\,The $\pi$ polarized light has a $\pi/2$ phase difference with the two circularly polarized lights. (iii)\,The Raman detuning $\Delta$ is far off resonance. 

For both the $F=1$ and $F=2$ systems, an effective $J_y$ Hamiltonian can be realized in the same way by changing the propagating directions of the Raman lasers from the $\hat{x}$ direction to the $\hat{y}$ direction.

\section{Dynamic decoupling}
In Section \uppercase\expandafter{\romannumeral3}, we give an effective rotation Hamiltonian by Raman process. Next, we utilize the effective Hamiltonian to construct a DD sequence. Through the dynamic evolution of the spin state, we give the optimal DD schemes for different atomic systems. 

Consider an atomic system with a total angular momentum $F$ (for the ground state of $^{87}\mathop{\rm Rb}$ $5 S_{1/2}$, $F=1$ or $2$) interacting with a leading magnetic field along $\hat{z}$ in the Earth-field range, we assume that the atomic spins are prepared in the $m_F = F$ state along the $\hat{z}$ direction ($J_z$) by an optical pumping (OP) field. The spin state is rotated to $-J_y$ by a $90^{\circ}x$ pulse to generate the coherence between Zeeman sublevels as quantum sensor or carrier of information. After that, the spin state begins rotating freely in the $ \hat x- \hat y$ plane because of Larmor precession. In presence of the NLZ effect and magnetic field gradient, the spin state begin diffusing, which broadens the linewidth of magnetic resonance and decreases the coherence time.
\subsection{NLZ cancellation}
To eliminate the $\textrm{NLZ}$ effect by $\textrm{DD}$, we choose the quadrupolar echo $\tau-90^{\circ}y-\tau$. 
With the $ H_{\rm eff} $ obtained in Section \uppercase\expandafter{\romannumeral3}, the spin rotation operations $90^{\circ}x$ and $90^{\circ}y$ can be realized by controlling the Rabi frequencies, detunings, propagating directions and pulse lengths of the Raman lasers.

For the $F=1$ system, we calculate the density matrix evolution under the quadrupolar echo $\tau-90^{\circ}y-\tau$ in detail in the Appendix B. In terms of the expression of $\rho(2\tau)$, the $\textrm{NLZ}$ effect still exists. If we choose an appropriate free evolution time $\tau$ such that $\Omega_L\tau=\pi$, $\rho(2\tau)$ turns into
\begin{equation}
	\begin{aligned}
		\rho(2\tau)=i\frac{1}{\sqrt{2}}\left( \begin{array}{ccc}   
			0 & 1 & 0\\  
			-1 & 0 & 1\\ 
			0 & -1 & 0\\
		\end{array}
		\right).
	\end{aligned}
\end{equation}
 To realize $\Omega_L\tau=\pi$, atomic spin can be drived by turning on the DD sequence with an optical rotation signal as the trigger (self-oscillating mode) \cite{higbie2006robust}. The final state is refocused to $-J_y$. An alternative approach is to carry out quadrature detection, in which only the components of $\rho(2\tau)$ irrelevant with respect to $\omega$ are detected \cite{antonijevic2003refocussing,levitt2013spin}. Thus, the $\textrm{NLZ}$ effect is removed by $90^{\circ}y$ pulses with arbitrary free evolution time $\tau$ as long as $\tau$ is much smaller than the NLZ dephasing time and much bigger than the pulse length.

In the $F=2$ system, due to more $\textrm{NLZ}$ splitting, the spin state dynamics under the quadrupolar echo is more complicated. To see the DD performance clearly, we present the density matrix evolution of the electron spin in the $J_y$ representation, as plotted in Fig\,4. Figs.\,4\,(a-d) show the real parts of the density matrix, and Figs.\,4\,(e-h) show the imaginary parts of the density matrix. In the $J_y$ representation, $-J_y$ means that atoms are fully populated in the $ m_F=-F$ state (Fig.\,4a,\,4e). During the free precession of state $-J_y$ in half of the Larmor precession period $\tau$ under the influence of the Larmor precession Hamiltonian $ \Omega_LJ_z $, the spin state rotates from $-J_y$ to $J_y$, as shown in Fig.\,4b. At the same time, the nondiagonal terms $\rho_{F,F-2}$ and $\rho_{F-2,F}$ occur because the $\textrm{NLZ}$ effect Hamiltonian $ \omega J_z^2$ brings $\Delta m_F=2$ transitions, as shown in Fig.\,4f. The produced coherence terms lead to a reduction in the atomic polarization. To eliminate the atomic decoherence effect, we apply a $90^{\circ}y$ pulse, which adds a $\pi$ phase to the terms $\rho_{F,F-2}$ and $\rho_{F-2,F}$ (Fig.\,4g) and keeps the target term $\rho_{F,F}$ unchanged (Fig.\,4c) (see Appendix C). Finally, the dephasing effect in the second free evolution $\tau$ counteracts the dephasing effect from the first free evolution $\tau$, which refocuses the final spin state to $-J_y$ (Fig.\,4d,\,4h).

\begin{figure}[htbp]
	\includegraphics[width=1\columnwidth]{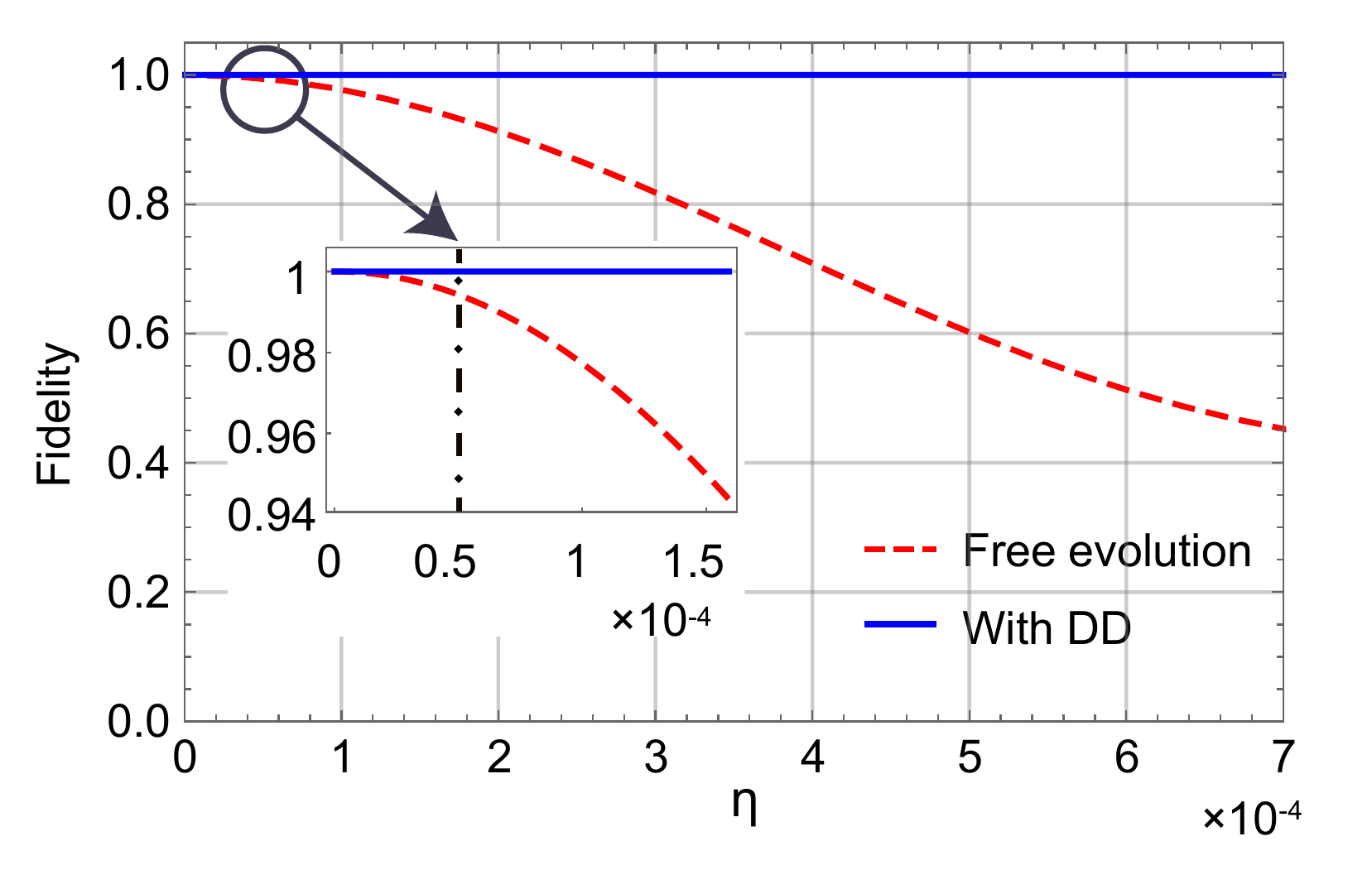} \caption{The state fidelity comparison between the cases free evolution and with DD applied after five cycles. $\eta$ is the ratio of the revival frequency $\omega$ and Larmor frequency $ \Omega_L $. Insert is the zoom in where real experimental parameters locate. Dashed line corresponds to the case of $^{87}\mathop{\rm Rb}$ atoms in the Earth-field range.} 
\end{figure}

Different from the $F = 1$ system, the density matrix of the $F=2$ system includes more nonzero terms induced by the NLZ effect. The $90^{\circ}y$ pulse cannot flip them all. In the $J_y$ representation, the $90^{\circ}y$ operation turns density matrix $\rho_{mn}$ to $\rho'_{mn}$ with $\rho'_{mn}=\exp[i(m+n)\pi/2]\rho_{mn}$, where m and n denote the magnetic quantum number (see Appendix C). There are 25 elements in the density matrix in total in the $F=2$ system, of which only $\rho_{2,0}$, $\rho_{0,2}$, $\rho_{-2,0}$ and $\rho_{0,-2}$ can be inverted by the $90^{\circ}y$ pulse, meaning that the quadrupolar echo cannot cancel out the $\textrm{NLZ}$ effect completely. The cancellation performance depends on the ratio $\eta$ of the revival frequency $\omega$ and Larmor frequency $ \Omega_L $. We define $\eta=\omega/\Omega_L$. The fidelity is introduced to evaluate the DD performance \cite{pokharel2018demonstration,F2} and formulated as $\mathcal{F}= {\rm Tr}|\sqrt{\sqrt{\rho_{1}'}\rho_{2}'\sqrt{\rho_{1}'}}|^2$  \cite{jeong2007quantum}. Here, $ \rho_{1}' $ is $ -J_y $ and $ \rho_{2}' $ is the state evolving for five DD cycles. Figure\,5 compares the fidelity against $ \eta $ in the cases with and without $\textrm{DD}$. From Fig.\,5, we find that the fidelity is maintained as high as $99.99\%$ in a wide range $ \eta <1\times10^{-3}$, where most atomic systems are located \cite{seltzer2008developments}. For example,
when $^{87}\mathop{\rm Rb}$ atoms operate in the $\textrm{Earth-field}$ range, the corresponding ratio $ \eta $ is $5.1\times10^{-5}$ (see Fig.\,5 inset).

\subsection{Magnetic field gradient cancellation}

A magnetic field gradient is inevitable in a natural environment. The phase perturbation due to an inhomogeneous magnetic field accumulates over time and causes a small spin orientation broadening along the target spin state, formulated as $\phi=(\Omega_L+\Delta\Omega_L)\tau$. First, this dephasing effect reduces the coherence time of the atomic electron spin. Second, for a long precession time and large phase perturbation, the condition $\Omega_L\tau=\pi$ cannot be achieved for all atoms. Hence, the quadrupolar echo cannot refocus the spin state perfectly, which compromises the performance of $\textrm{NLZ}$ effect cancellation.

Dephasing due to an inhomogeneous magnetic field can be largely alleviated by using the $\textrm{Hahn}$ echo  \cite{rui2015operating,hahn1950spin}, where a $180^{\circ}y$ pulse is applied during the free evolution interval. In our scheme, the $180^{\circ}y$ pulse is realized by Raman scattering.
\begin{figure}[htbp]
	\includegraphics[width=1\columnwidth]{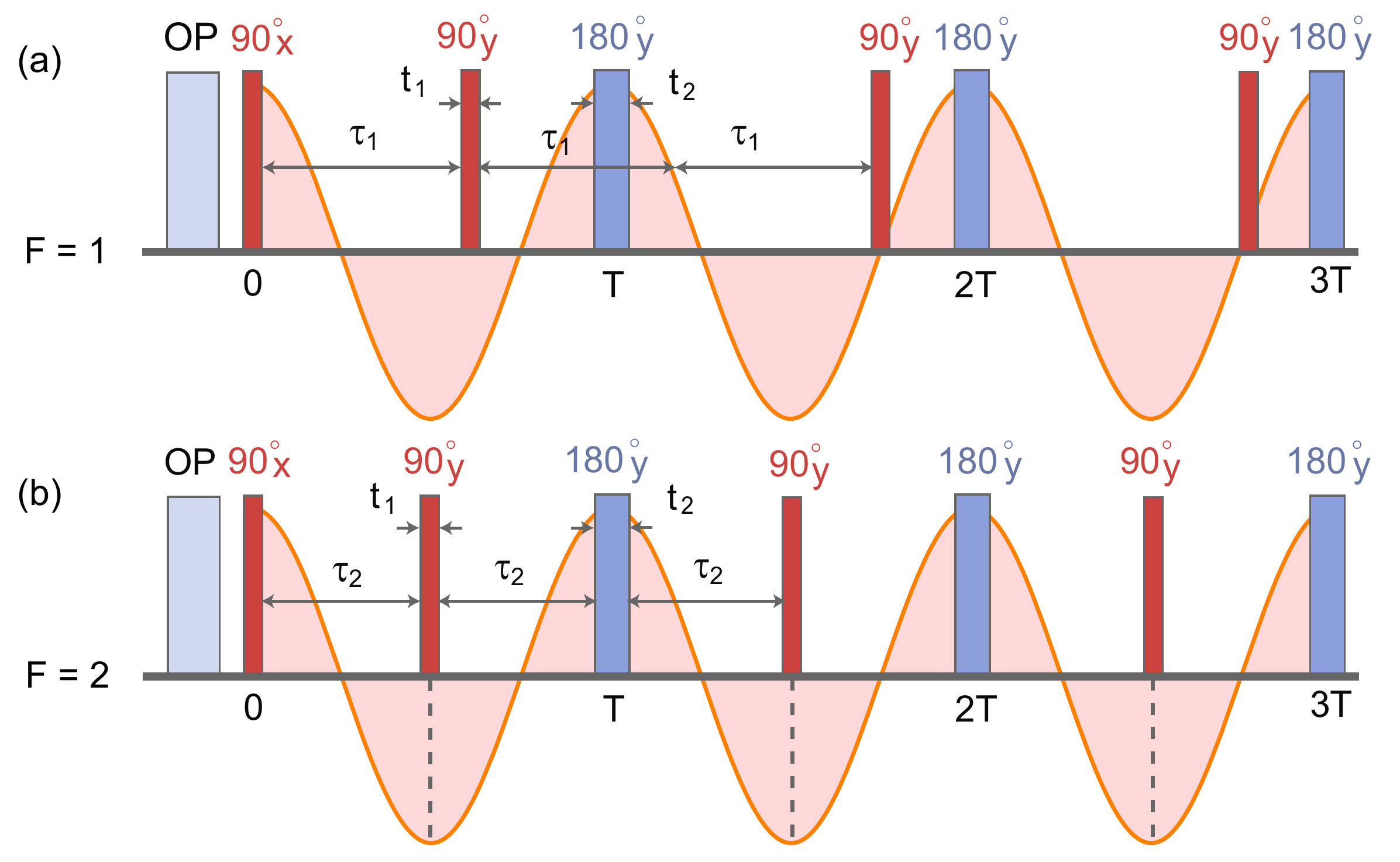} \caption{DD control sequence by inserting Hahn echo $\tau-180^{\circ}y-\tau$ into quadrupolar echo $\tau-90^{\circ}y-\tau$ for (a) $ F=1 $ system. (b) $ F=2 $ system. The curved line represents projection of the evolved spin state to $J_y$ in the Larmor precession process. The Raman pulses are all square waves. Larmor period $ T $ is 28.5\,$\mu$s. The free evolution time $\tau_1$ is adjustable from about 14.3\,$\mu$s to about 5.59\,ms and $\tau_2$ is 14.3\,$\mu$s. $90^{\circ}y$ pulse length $ t_1 $is 2.85\,$\mu$s. $180^{\circ}y$ pulse length $ t_2 $ is 5.7\,$\mu$s. } 
\end{figure}	

 To remove the magnetic field gradient altogether, we apply a $180^{\circ}y$ pulse at the end of each Larmor cycle, as shown in Fig.6. Note that the $180^{\circ}y$ pulses have no influence on the $\textrm{NLZ}$ effect cancellation (see Appendix C). Figure 6 gives the specific DD control schemes. The Larmor frequency for $^{87}\mathop{\rm Rb}$ in the Earth's magnetic field is $\Omega_L=2\pi\times350$\,kHz. The corresponding Larmor period $ T $ is 28.5\,$\mu$s. For $ F=1 $ system, the free evolution time $\tau_1$ is arbitrary as long as it is smaller than the NLZ dephasing time and much bigger than the pulse length using quadrature detection. Based on the revival frequency $\omega=2\pi\times17.9$\,kHz, $\tau_1$ is adjustable from about 14.3\,$\mu$s to about 5.59\,ms. For $ F=2 $ system, the free evolution time $\tau_2$ is half of the Larmor period 14.3\,$\mu$s. To get a fast rotation operation, the length of square Raman pulse can be set as $ 2.85\,\mu$s for $ 90^{\circ} $ rotation and $ 5.7\,\mu$s for $ 180^{\circ} $ rotation. Thus, the effective Rabi frequency $\Omega_1\Omega_2/(24\Delta)$ is $2\pi\times0.875$\,MHz. For a Raman detuning 4\,GHz, Rabi frequencies of the three polarized Raman lights can be set as $2\pi\times$290\,MHz equally.

\section{Conclusion and discussion}
We demonstrate an all-optical $\textrm{DD}$ scheme to suppress the $\textrm{NLZ}$ effect and magnetic field gradient at the same time. We design an all-optical control sequence by inserting Hahn echo $\tau-180^{\circ}y-\tau$ into quadrupolar echo $\tau-90^{\circ}y-\tau$ via Raman process. The NLZ effect can be canceled out by the quadrupolar echo. There are two ways to set the free evolution time $\tau$. One is to set $\tau$ to half of the Larmor precession period, which can be realized by self-oscillating. The other is to use arbitrary $\tau$ with quadrature detection as long as $\tau$ is much smaller than the NLZ dephasing time and much bigger than the pulse length. For the $F=1$ system, the $\textrm{NLZ}$ effect can be completely removed by both methods. For the $F=2$ system, the $90^{\circ}y$ pulses are added at every half time of the Larmor precession period. Due to the greater amount of $\textrm{NLZ}$ splitting, the $\textrm{NLZ}$ effect cannot be completely canceled out, and the decoupling results rely on the ratio $ \eta $ of the revival frequency to the Larmor precession frequency. By analyzing the fidelity of the atomic spin state with and without $\textrm{DD}$, we find that the fidelity is maintained as high as $99.99\%$ in a wide range $\eta <1\times10^{-3}$ after five $\textrm{DD}$ cycles. The magnetic field gradient can be completely eliminated by $\textrm{Hahn}$ echo.

For $^{87}\mathop{\rm Rb}$ in the $\textrm{Earth-field}$ range, the signal splitting due to the NLZ effect is $ 2\pi\times35.8$\,Hz for $ F=1 $ system and $ 2\pi\times 71.6$\,Hz for $ F=2 $ system. The typical relative inhomogeneity of magnetic field is about $ 10^{-4} $ \cite{magnetic}, leading to the signal broadening of $ 2\pi\times35$\,Hz. As a consequence, the coherence time is 27.9\,ms for $ F=1 $ system and 14\,ms for $ F=2 $ system. After removing the NLZ effect and magnetic field gradient by DD, the coherence time is only limited by the electron spin-destruction collisions with the cell walls, residual relaxation due to collisions with the reservoir, and alkali-metal–alkali-metal spin-exchange collisions, which reaches 60\,s \cite{77s}. The sensitivity of atomic magnetometer can be improved by 46.4 times for $ F=1 $ system and 65.5 times for $ F=2 $ with DD applied.

The Larmor frequency for $^{87}\mathop{\rm Rb}$ in the Earth's magnetic field is $\Omega_L=2\pi\times350$\,kHz. 
To get a fast rotation operation, the length of square Raman pulse can be set as $ 2.85\,\mu$s for $ 90^{\circ} $ rotation and $ 5.7\,\mu$s for $ 180^{\circ} $ rotation. The effective Rabi frequency is $2\pi\times0.875$\,MHz. For a Raman detuning 4\,GHz, Rabi frequencies of the three polarized Raman lights can be set as $2\pi\times$290\,MHz equally. Our pulsed $\textrm{DD}$ scheme suppresses the power broadening significantly compared with the continuous spin locking approach. Furthermore, this all-optical $\textrm{DD}$ does not bring crosstalk between adjacent samples, which is essential for multisensor detection. It has promising prospects for practical application in quantum sensing and quantum information.
\section*{Acknowledgments}
The authors would like to thank Min Jiang from the University of Science and Technology of China for useful discussions. The authors acknowledge financial support from the National Key Research and Development Program of China (2016YFA0302001); the National Natural Science Foundation of China (11874152, 11654005); the Fundamental Research Funds for the Central Universities; the Shanghai Municipal Science and Technology Major Project (2019SHZDZX01); and the Fellowship of China Postdoctoral Science Foundation (2020TQ0193); W. Z. also acknowledges additional support from the Shanghai talent program.
\appendix
\section{Adiabatic Approximation}
When multilevel atoms interact with optical fields, the block matrix form of the interaction Hamiltonian can be written as
\begin{equation}
	\begin{aligned}
		H=\left(                 
		\begin{array}{cc}   
			H_{gg} & H_{ge}\\  
			H_{eg} & H_{ee}\\
		\end{array}
		\right).
	\end{aligned}
\end{equation}
The corresponding block density matrix is
\begin{equation}
	\begin{aligned}
		\rho=\left(                 
		\begin{array}{cc}   
			\rho_{gg} & \rho_{ge}\\  
			\rho_{eg} & \rho_{ee}\\
		\end{array}
		\right).
	\end{aligned}
\end{equation}

Substituting these two expressions into the Liouville equation $id\rho/dt=[H,\rho]$, we have
\begin{subequations} \label{e:1234}
	\begin{align}
		\label{12345}	
		i\frac{d\rho_{gg}}{dt}&=H_{gg}\rho_{gg}+H_{ge}\rho_{eg}-\rho_{gg}H_{gg}-\rho_{ge}H_{eg},\\    
		\label{12345}     
		i\frac{d\rho_{ge}}{dt}&=H_{gg}\rho_{ge}+H_{ge}\rho_{ee}-\rho_{gg}H_{ge}-\rho_{ge}H_{ee},\\	 
		\label{12345}
		i\frac{d\rho_{eg}}{dt}&=H_{eg}\rho_{gg}+H_{ee}\rho_{eg}-\rho_{eg}H_{gg}-\rho_{ee}H_{eg},\\	 
		\label{12345}
		i\frac{d\rho_{ee}}{dt}&=H_{eg}\rho_{ge}+H_{ee}\rho_{ee}-\rho_{eg}H_{ge}-\rho_{ee}H_{ee}.
	\end{align}
\end{subequations}

Usually, the ground-state Zeeman splitting is hundreds of kHz, while the Raman detuning $\Delta$ is several GHz. Thus, compared with $H_{ee}$, $H_{gg}$ is negligible. In addition, the Raman detuning $\Delta$ is much larger than the natural linewidth of the excited energy level $5P_{1/2}$. Therefore, we can carry out an adiabatic approximation for the excited state, that is, $d\rho_{ge}/dt=0$, $d\rho_{eg}/dt=0$. Additionally, in the Raman scattering process, the number of excited atoms is far less than the number of unexcited atoms. Thus, we can simply set $\rho_{ee}=0$. With all the assumptions above, Eq.\,(A3) is reduced to
\begin{equation}
	\begin{aligned}
		0&=-\rho_{gg}H_{ge}-\rho_{ge}H_{ee},\\
		0&=H_{eg}\rho_{gg}+H_{ee}\rho_{eg}.\\
	\end{aligned}
\end{equation}

Then, we have
\begin{equation}
	\begin{aligned}
		\rho_{ge}&=-\rho_{gg}H_{ge}H_{ee}^{-1},\\
		\rho_{eg}&=-H_{ee}^{-1}H_{eg}\rho_{gg}.\\
	\end{aligned}
\end{equation}

Substituting Eq.\,(A5) into Eq.\,(A3), we end up with
\begin{equation}
	\begin{aligned}
		i\frac{d\rho_{gg}}{dt}=&H_{gg}\rho_{gg}-H_{ge}H_{ee}^{-1}H_{eg}\rho_{gg}-\\
		&\rho_{gg}H_{gg}+\rho_{gg}H_{ge}H_{ee}^{-1}H_{eg}\\
		=&[H_{gg}+H_{eff},\rho_{gg}].
	\end{aligned}
\end{equation}

Furthermore, we obtain
\begin{equation}
	H_{eff}=-H_{ge}H_{ee}^{-1}H_{eg}.
\end{equation}

With the help of the ADM package in Mathematica \cite{ADM}, we can write out the specific efficient Hamiltonian on the basis of the Zeeman sublevels.

\section{State evolution with the quadrupolar echo}
For the $F=1$ system, we calculate the state evolution under the quadrupolar echo $\tau-90^{\circ}y-\tau$. Initially, the atoms are populated in the $J_z$ state. The density matrix after a $90^{\circ}x$ pulse evolves into
\begin{equation}
	\begin{aligned}
		J_z \stackrel{90^{\circ}x}{\longrightarrow}-J_y=\rho(0).
	\end{aligned}
\end{equation}
Free precession of this state for a period $\tau$ under the influence of the Larmor precession Hamiltonian $ \Omega_LJ_z $ and the NLZ effect Hamiltonian $ \omega J_z^2 $ produces the following transformations:
\begin{equation}
	\begin{aligned}
		\stackrel{\Omega_L J_z\tau}{\longrightarrow}-&[J_y\cos(\Omega_L\tau)-J_x\sin(\Omega_L\tau)],\\
		\stackrel{\omega J_z^2\tau}{\longrightarrow}-&\left\{ \left[J_y \cos(\Omega_L\tau)-i C_y \sin(\omega\tau)\right]\cos(\Omega_L\tau)-\right.\\
		&\left.[J_x \cos(\Omega_L\tau)-i C_x \sin(\omega\tau)] \sin(\Omega_L\tau)\right\}.
	\end{aligned}
\end{equation}
Since $ \Omega_LJ_z $ commutates with $ \omega J_z^2 $, the order in which we perform the transformations in Eq.\,(B2) is irrelevant.	

Then, a $ 90^{\circ}y $ pulse is applied to the final state in $\textrm{Eq}$.\,(B2), yielding
\begin{equation}
	\begin{aligned} 
		\stackrel{90^{\circ}y}{\longrightarrow}-&\left\{[J_y \cos(\omega\tau)+i C_y \sin(\omega\tau)]\cos(\Omega_L\tau)-\right.\\
		&\left.[-J_z \cos(\omega\tau)-i C_q \sin(\omega\tau)]\sin(\Omega_L\tau)\right\}.
	\end{aligned}
\end{equation}

After the $ 90^{\circ}y $ pulse, an echo is formed following a second period $ \tau $ of free precession:
\begin{equation}
	\begin{aligned} 
		\stackrel{\Omega_L J_z\tau}{\longrightarrow}-&\left\{\left\{[J_y\cos(\Omega_L\tau)-J_x\sin(\Omega_L\tau)]\cos(\omega\tau)\right.\right.\\
		&+\frac{1}{\sqrt{2}}[C_p\cos(\Omega_L\tau)-i C_a \sin(\Omega_L\tau)]\\
		&\left.\sin(\omega\tau)\right\}\cos(\Omega_L\tau)\left\{-J_z \cos(\omega\tau)-i[C_q \right.\\
		&\left.\cos(2\Omega_L\tau)-i C_b \sin(2\Omega_L\tau)]\sin(\omega\tau)\right\}\\
		&\left.\sin(\Omega_L\tau)\right\},
	\end{aligned}
\end{equation}
\begin{equation}
	\begin{aligned} 
		\stackrel{\omega J_z^2\tau}{\longrightarrow}-&\left\{\left\{[J_y \cos(\omega\tau)-i C_y \sin(\omega\tau)]\cos(\Omega_L\tau)\right.\right.\\
		&-[J_x \cos(\omega\tau)-i C_x \sin(\omega\tau)]\cos(\omega\tau)\\
		&+\frac{1}{\sqrt{2}}\left\{[C_p \cos(\omega\tau)-i C_i \sin(\omega\tau)]\right.\\
		&\cos(\Omega_L\tau)-i[C_a \cos(\omega\tau)-i\sqrt{2}J_x \\
		&\left.\left.\sin(\omega\tau)]\sin(\Omega_L\tau)\right\}\sin(\omega\tau)\right\}\cos(\Omega_L\tau)- \\
		&\left\{\left\{-J_z\cos(\omega\tau)-i[C_q\cos(2\Omega_L\tau)-i C_b \right.\right.\\
		&\left.\left.\left.\sin(2\Omega_L\tau)]\sin(\omega\tau)\right\}\right\}\sin(\Omega_L\tau)\right\}=\rho(2\tau),
	\end{aligned}
\end{equation}
where 

$C_x=\frac{1}{\sqrt{2}}\left(                 
\begin{array}{ccc}   
	0 & 1 & 0\\  
	-1 & 0 & -1\\ 
	0 & 1 & 0\\
\end{array}
\right)$,
$C_y=i\frac{1}{\sqrt{2}}\left(                 
\begin{array}{ccc}   
	0 & -1 & 0\\  
	-1 & 0 & 1\\ 
	0 & 1 & 0\\
\end{array}
\right)$,		

$C_n=\frac{1}{\sqrt{2}}\left(                 
\begin{array}{ccc}   
	0 & 0 & 1\\  
	0 & 0 & 0\\ 
	-1 & 0 & 0\\
\end{array}
\right)$,
$C_i=\left(                 
\begin{array}{ccc}   
	0 & 1 & 0\\  
	-1 & 0 & 1\\ 
	0 & -1 & 0\\
\end{array}
\right)$,	

$C_a=\left(                 
\begin{array}{ccc}   
	0 & 1 & 0\\  
	-1 & 0 & -1\\ 
	0 & 1 & 0\\
\end{array}
\right)$,
$C_p=i\sqrt{2}C_y,C_q=\sqrt{2}C_n$.	

With these substitutions, the matrix form of $ \rho(2\tau) $ at the end of the second free precession period is
\begin{widetext}
	\begin{equation}
		\begin{split}
			\rho(2\tau)=\left(                 
			\begin{array}{ccc}   
				-\cos(w\tau)\sin(\Omega_L \tau) & \frac{\cos(\Omega_L \tau)(i \cos(\Omega_L \tau)+\sin(\Omega \tau))}{\sqrt{2}} & -i e^{-2i \Omega_L\tau} \sin(w\tau)\sin(\Omega_L \tau)\\  
				\frac{-i (1+e^{-2 i \Omega_L \tau})}{2\sqrt{2}} & 0 & \frac{\cos(\Omega_L \tau)(i \cos(\Omega_L \tau)+\sin(\Omega_L \tau))}{\sqrt{2}}\\ 
				i e^{-2 i \Omega\tau}\sin(w\tau)\sin(\Omega_L \tau) & \frac{i (1+e^{-2 i \Omega_L \tau})}{2\sqrt{2}} & \cos(w\tau)\sin(\Omega_L \tau)\\			
			\end{array}
			\right).
		\end{split}
	\end{equation}
\end{widetext}
\section{Spin rotation in the $J_y$ representation}
In the $J_y$ representation, the matrix form of the $90^{\circ}y$ operation on the basis of the $F=2$ Zeeman sublevels is
\begin{equation}
	\begin{aligned}
		{\rm rot}_{90^{\circ}y}=\left( \begin{array}{ccccc}   
			-1 & 0 & 0 & 0 & 0\\  
			0 & i & 0 & 0 & 0\\ 
			0 & 0 & 1 & 0 & 0\\
			0 & 0 & 0 & i & 0\\
			0 & 0 & 0 & 0 & -1
		\end{array}
		\right).
	\end{aligned}
\end{equation}

With the $90^{\circ}y$ pulse applied, an arbitrary density matrix $\rho$ 
\begin{equation}
	\begin{aligned}
		\rho=\left( \begin{array}{ccccc}   
			\rho_{2,2} & \rho_{2,1} & \rho_{2,0} & \rho_{2,-1} & \rho_{2,-2}\\  
			\rho_{1,2} & \rho_{1,1} & \rho_{1,0} & \rho_{1,-1} & \rho_{1,-2}\\ 
			\rho_{0,2} & \rho_{0,1} & \rho_{0,0} & \rho_{0,-1} & \rho_{0,-2}\\
			\rho_{-1,2} & \rho_{-1,1} & \rho_{-1,0} & \rho_{-1,-1} & \rho_{-1,-2}\\
			\rho_{-2,2} & \rho_{-2,1} & \rho_{-2,0} & \rho_{-2,-1} & \rho_{-2,-2}\\
		\end{array}
		\right)
	\end{aligned}
\end{equation}
is transformed into
\begin{equation}
	\begin{aligned}
		\rho'=
		\left( \begin{array}{ccccc}   
			\rho_{2,2} & i\rho_{2,1} & -\rho_{2,0} & -i\rho_{2,-1} & \rho_{2,-2}\\  
			-i\rho_{1,2} & \rho_{1,1} & i\rho_{1,0} & -\rho_{1,-1} & -i\rho_{1,-2}\\ 
			-\rho_{0,2} & -i\rho_{0,1} & \rho_{0,0} & i\rho_{0,-1} & -\rho_{0,-2}\\
			i\rho_{-1,2} & -\rho_{-1,1} & -i\rho_{-1,0} & \rho_{-1,-1} & i\rho_{-1,-2}\\
			\rho_{-2,2} & i\rho_{-2,1} & -\rho_{-2,0} & -i\rho_{-2,-1} & \rho_{-2,-2}\\
		\end{array}
		\right),
	\end{aligned}
\end{equation}
with the unitary transform $\rho'={\rm rot}_{90^{\circ}y}\rho {\rm rot}_{90^{\circ}y}^\dagger$.

A uniform expression for the density matrix elements is
\begin{equation}
	\begin{aligned}
		\rho'_{mn}=\exp(i\frac{m+n}{2}\pi)\rho_{mn},
	\end{aligned}
\end{equation}
where $m$ and $n$ are magnetic quantum numbers.

From Eq.\,(C4), we can see that the density matrix obtains a $\pi$ phase for the elements $\rho_{2,0}$, $\rho_{0,2}$, $\rho_{-2,0}$ and $\rho_{0,-2}$ while keeping the diagonal elements unchanged.

The $180^{\circ}y$ operation is equivalent to two  $90^{\circ}y$ operations. A uniform expression for density matrix elements under the $180^{\circ}y$ operation is
\begin{equation}
	\begin{aligned}
		\rho'_{mn}=\exp[i(m+n)\pi]\rho_{mn}.
	\end{aligned}
\end{equation} 

\end{document}